\documentclass[sigconf, natbib=false]{acmart}
\newcommand{\myparagraph}[1]{\noindent \textbf{#1 }}

\copyrightyear{2026}
\acmYear{2026}
\setcopyright{cc}
\setcctype{by}
\acmConference[ICSE-Companion '26]{2026 IEEE/ACM 48th International Conference on Software Engineering}{April 12--18, 2026}{Rio de Janeiro, Brazil}
\acmBooktitle{2026 IEEE/ACM 48th International Conference on Software Engineering (ICSE-Companion '26), April 12--18, 2026, Rio de Janeiro, Brazil}
\acmDOI{10.1145/3774748.3787623}
\acmISBN{979-8-4007-2296-7/2026/04}

\title[A Multi-Modal Robotic Framework for Physical GUI Interaction Testing]{A Multi-Modal Robotic Framework for Physical GUI Interaction Testing}

\author{Davi Freitas}
\orcid{0009-0000-1419-5206}
\affiliation{
  \institution{Federal University of Pernambuco (UFPE)}
  \country{Brazil}
}
\email{dsf3@cin.ufpe.br}

\begin{document}

\begin{abstract}
Automated testing of mobile applications remains challenging when realistic user interactions must be reproduced and GUI behavior verified without relying on invasive mechanisms such as instrumentation or ADB-level control. Existing approaches, including DLD and R-DLD, detect data loss triggered by activity restarts but still depend on system-level access, which limits their applicability on real devices.

This doctoral research proposes RFNIT, a multi-modal, fully non-invasive robotic framework for physical GUI interaction testing on smartphones. RFNIT supports end-to-end automation by perceiving the device state through external sensing and executing physical interactions, enabling human-like actions while preserving reproducibility and execution accuracy.

The framework integrates a vision-based perception module, an LLM-driven reasoning agent, and a robotic actuator capable of performing touch, swipe, and typing gestures. Multimodal data, including external camera images and OCR-extracted text, are combined to identify GUI elements and guide subsequent testing decisions.

Preliminary results indicate that RFNIT accurately recognizes buttons, text fields, and virtual keyboards from external camera input, achieving over 85\% visual detection accuracy. Ongoing and future work will focus on improving robustness, repeatability, and coverage of GUI states during long-running testing sessions.

\end{abstract}

\keywords{Non-invasive testing, robotics testing, computer vision, mobile applications, GUI testing}

\maketitle

\section{Introduction}

Automated testing is essential to ensure software quality and reliability. However, as applications move beyond smartphones to platforms such as automotive dashboards, industrial controllers, IoT devices, and gaming consoles, traditional frameworks like Appium~\cite{appium}, Espresso~\cite{espresso}, and UIAutomator~\cite{UIAutomator} become increasingly limited. These tools rely on invasive APIs or instrumentation, which are often unavailable in proprietary or safety-critical environments, forcing many domains to depend on costly, error-prone, and poorly reproducible manual testing.

Non-invasive robotic testing has emerged as an alternative to validate mobile applications without instrumentation, using external sensors and actuators to simulate human interactions. Approaches such as RoScript~\cite{roscript, AxizBlackBoxAutomation} demonstrate the feasibility of physical touch automation; however, by keeping the device static on a bench while a stylus executes gestures, these systems remain limited to two-dimensional touchscreen interactions. As a result, motion-dependent sensors, physical buttons, and complex gestures—such as orientation changes, “double chop for flashlight,” “twist for camera,” or NFC-based interactions—remain out of scope for non-invasive automation.

\begin{figure}[h]
\centering
    \includegraphics[width=1\columnwidth]{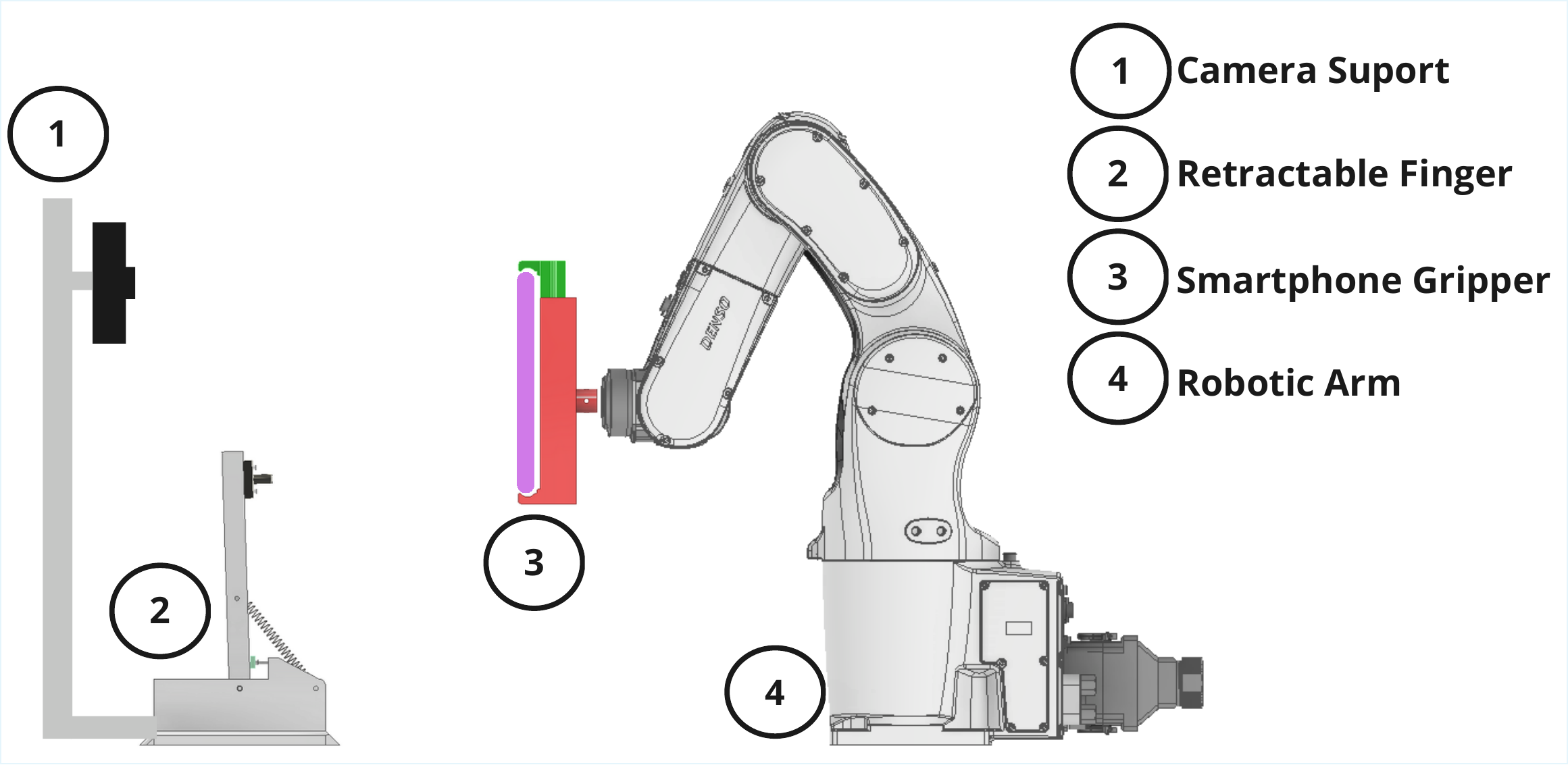}    
    \caption{Robotic structure of RFNIT.}
    \label{fig:robo_desenho}
\end{figure}

To overcome this gap, RFNIT adopts an inverted configuration: the smartphone is mounted directly on the robot, while the stylus and external camera remain fixed in the environment (Figure~\ref{fig:robo_desenho}). This design enables the robot to physically manipulate the device, perform controlled rotations, approximate NFC cards, and trigger advanced motion-based gestures while still supporting precise touch input. This expands automated testing from simple 2D screen interactions to a broader, multimodal physical interaction space.

Preliminary steps of this research were explored in earlier works. An initial RFNIT prototype was presented as a poster at FSE 2024 \cite{rfnit_fse2024}, validating the feasibility of robotic-based, less invasive testing on real smartphones. In parallel, a study at SBES 2024~\cite{rdld_sbes2024} investigated robotic-supported data loss detection, providing early evidence that physical device manipulation can reduce test invasiveness. These insights directly motivated the design of the fully non-invasive and multimodal RFNIT framework proposed in this doctoral research.

Building on these insights, this doctoral research proposes RFNIT (Robotic Framework for Non-Invasive Testing), a multi-modal robotic framework that integrates robotics, computer vision, and automated reasoning to execute high-fidelity physical GUI interaction testing on heterogeneous devices. RFNIT provides a two-layer modular architecture: (i) a physical layer that controls actuation and sensing, and (ii) a software layer that abstracts low-level control and exposes a high-level intent-driven API for defining test behaviors.

The key differentiators of RFNIT are:

\myparagraph{Multimodal support:} execution of touchscreen gestures, physical button presses, controlled device rotations, and advanced gestures such as double-tap and drag, enabling interaction with both on-screen and hardware-dependent features;

\myparagraph{Real-time adaptation:} visual- and OCR-based perception supports responsive interaction, on-the-fly correction of execution deviations, and recovery from unexpected interface changes;

\myparagraph{Device-agnostic design:} applicability to Android, iOS, and embedded platforms without requiring instrumentation, system-level access, or modifications to the application under test;

\myparagraph{Integration with testing ecosystems:} compatibility with Pytest for defining automated oracles, structuring test cases, and integrating RFNIT into continuous testing and CI pipelines.

RFNIT aims to push the boundaries of non-invasive test automation by providing a framework capable of validating not only the functional correctness of applications but also the fidelity, robustness, and user experience of physical interactions.

\section{Research Problem and Objectives}

Despite the progress in GUI testing automation, a major gap persists between invasive software-level testing and non-invasive physical testing. Invasive methods such as ADB- or instrumentation-based testing can achieve fine-grained control of app states but fail to operate in realistic user conditions. Conversely, robotic systems that execute touch events on real devices are constrained by limited perception and reasoning capabilities. Bridging this gap requires a unified framework that enables robots to interpret and act upon the visual state of the device, autonomously and intelligently.

The overarching research problem of this thesis can be summarized as follows:

\begin{quote}
\textit{How can we design a robotic testing framework that performs reliable, autonomous, and human-like GUI interactions on real smartphones without requiring invasive system access?}
\end{quote}

To address this problem, the research is guided by the following objectives:

\myparagraph{O1 – Develop a non-invasive architecture for automated GUI testing:} design a framework that integrates vision-based perception, LLM-driven reasoning, and robotic actuation to perform end-to-end interactions on real devices.  

\myparagraph{O2 – Enable multimodal understanding of interface states:} combine external camera images, and OCR-extracted text to accurately identify interactive components and infer app states. 

\myparagraph{O3 – Achieve adaptive testing strategies:} implement a decision-making mechanism that allows the framework to select the next interaction step based on its perception and testing goals.  

\myparagraph{O4 – Evaluate the approach in realistic environments:} validate RFNIT’s performance through experiments involving real Android applications, measuring coverage, repeatability, and robustness during long testing sessions.

By pursuing these objectives, RFNIT aims to extend the current boundaries of mobile testing automation, bringing together the precision of robotic actuation and the flexibility of AI-driven reasoning. The resulting framework seeks to advance the understanding of how perception, cognition, and physical interaction can coexist in software testing tools for mobile platforms.

\section{Proposed Approach: The RFNIT Framework}

\begin{figure}[t]
    \centering
    \includegraphics[width=0.78\linewidth]{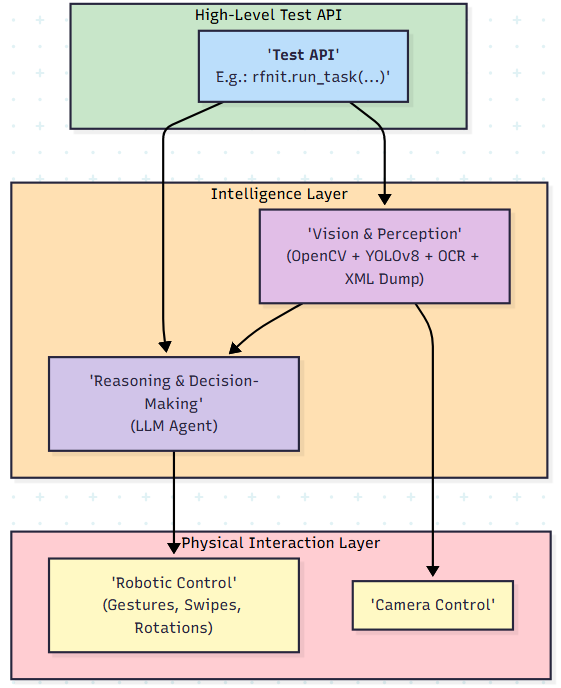}
    \caption{RFNIT architecture integrating perception, reasoning, and physical control for non-invasive testing.}
    \label{fig:rfnit_architecture}
\end{figure}

The proposed \textbf{RFNIT} integrates three complementary modalities—vision, reasoning, and control—into a unified non-invasive testing architecture (Figure~\ref{fig:rfnit_architecture}).

\subsection{Vision and Perception}
An external camera (Logitech BRIO 4K) captures real-time images of the smartphone screen. The visual processing pipeline combines OpenCV-based preprocessing for perspective correction and brightness normalization, layout extraction from Android XML dumps, object detection using YOLOv8 trained on GUI datasets, and text recognition via PaddleOCR. The result is a multimodal representation of the interface that integrates visual and semantic information about buttons, text fields, and interactive regions.

\subsection{Reasoning and Decision-Making}
An LLM acts as a decision-making agent that interprets the visual state of the interface and the testing task to select the next interaction. The agent performs goal-driven exploration guided by natural language instructions, reasons about application state transitions and missing elements to identify potential data loss or UI faults, and continuously adapts its strategy based on feedback from prior executions.

\subsection{Physical Interaction}
The robot executes the selected action physically on the smartphone through a controlled actuation mechanism. Each touch, swipe, or typing command is translated into precise servo or actuator control signals. Visual feedback is continuously used to verify whether the intended action produced the expected interface state transition, enabling a closed-loop perception–action feedback cycle that improves robustness and execution accuracy.

\section{Preliminary Results}

A functional RFNIT prototype was implemented using Python, OpenCV, and the Ultralytics YOLOv8 framework. Experiments were conducted on five Android applications (e.g., messaging, weather, and shopping apps), using an external Logitech BRIO 4K camera operating under natural ambient lighting conditions. The system successfully identified interactive components and executed physical tap actions through a robotic actuator with an average precision of 2.0 mm.

Visual detection achieved 91\% accuracy (mAP@0.5) when bounding boxes were compared against ADB-based ground truth. OCR text recognition exceeded 95\% accuracy under varying natural lighting, confirming the robustness of the perception pipeline in realistic, non-controlled scenarios. Furthermore, because the smartphone is rigidly attached to the robotic arm, RFNIT can rotate the device and press its side buttons (volume up/down and power) by aligning the device with a fixed actuator. This enables automated testing of sensor-based and hardware-button interactions that are inaccessible to traditional static robotic setups. These results demonstrate the feasibility of end-to-end multimodal perception and physical interaction without instrumentation.

\section{Evaluation Plan}

The next phase of this research will conduct a comprehensive evaluation of RFNIT along the following dimensions:

\myparagraph{Coverage and efficiency:} Quantitatively compare state coverage, execution time, and the number of unique interaction transitions against established baselines, including DLD and DroidBot.

\myparagraph{Realism and usability:} Perform controlled user studies to assess the degree to which robotic actions align with human interaction patterns, focusing on perceived naturalness, gesture fidelity, and operational intuitiveness.

\myparagraph{Reproducibility and robustness:} Execute repeated test sessions under varying environmental and device conditions. Metrics such as visual detection precision, action latency, and physical alignment accuracy will be systematically measured to characterize stability and performance.

Overall, this evaluation aims to rigorously demonstrate the viability of RFNIT as a general-purpose, multimodal, and fully non-invasive robotic testing framework for real mobile devices.

\section{Challenges and Future Work}

Despite promising initial results, several challenges remain:

\myparagraph{Latency and synchronization:} Aligning perception, reasoning, and actuation in real time remains complex, especially as gesture complexity increases.

\myparagraph{Visual occlusion and reflection:} External cameras are sensitive to glare and variability in ambient light, which can interfere with computer vision pipelines.

\myparagraph{Reasoning reliability:} LLMs may produce inconsistent decisions when faced with ambiguous or partially occluded interfaces.

\myparagraph{Scalability:} Extending testing to multiple devices and concurrent executions requires sophisticated coordination strategies.


To extend the platform’s physical capabilities, we are developing a new robotic gripper that supports rapid smartphone replacement and magnetic USB coupling for automatic power connection. In parallel, we are designing a dynamic grip reconfiguration mechanism that allows the robot to change the device’s orientation or handedness during testing, enabling inversion, rotation, and repositioning to support complex gestures and access to side buttons.

\section{Expected Contributions}

This research is expected to make the following contributions:

    \myparagraph{A fully non-invasive robotic testing framework} for multimodal GUI interaction on physical mobile devices, enabling external vision--based perception, reasoning-driven action selection, and precise robotic execution without requiring instrumentation or system-level access.

    \myparagraph{A publicly available, large-scale dataset} composed of synchronized camera captures, supporting research on visual understanding, GUI element detection, and multimodal test generation.

    \myparagraph{Empirical evidence and methodological insights} on how large reasoning models can be integrated with physical testing workflows, including analysis of decision reliability, perception--action coupling, and failure modes in real-world scenarios.

    \myparagraph{A conceptual and practical foundation for hybrid AI--robotic testing methodologies}, bridging software testing, mobile automation, computer vision, and robotic manipulation to support future research on autonomous, embodied testing agents.

\bibliographystyle{ACM-Reference-Format}
\bibliography{software.bib}

\end{document}